\documentclass[
    ,final            
  ]
  {aipproc}

\layoutstyle{6x9}

\newcommand{\be}{\begin{equation}}
\newcommand{\ee}{\end{equation}}
\newcommand{\ba}{\begin{array}{c}}
\newcommand{\ea}{\end{array}}
\newcommand{\bqa}{\begin{eqnarray}}
\newcommand{\eqa}{\end{eqnarray}}
\begin{document}

\title{Dynamical Properties of the $\sigma$ Meson\footnote{Talk given at \textit{Workshop
        on Scalar Mesons and Related Topics} honoring the 70th birthday of Michael Scadron,
      Feb. 11-16 2008,   Lisbon, Portugal}}



\author{H.~Q.~Zheng}{
  address={Department of Physics, Peking University, Beijing 100871, P.~R.~China}
}

\begin{abstract}
 Studies on the dynamical properties of the $\sigma$ meson are
reviewed and discussed. The important role of fundamental principles
such as analyticity, unitarity and crossing symmetry played in the
studies are stressed.
\end{abstract}

\maketitle

Remarkable progress has been made in recent few years in revealing
unambiguously  the existence of the light and broad resonance
$f_0(600)$,~\cite{review1} for which dispersive analyses play a
crucial role.~\cite{dis1,dis2} Nevertheless there still remains
difficulties in understanding  the nature of the $f_0(600)$ or the
$\sigma$ meson   at the fundamental (i.e., QCD) level, though
efforts are  started to make from lattice approach.~\cite{lattice}

There exist, however, at phenomenological level, many investigations
on the dynamical property of the $\sigma$ meson. Most of these
studies are based upon $\sigma$-like  models.~\cite{Lsigma} There
are also attempts trying to interpret the $f_0(600)$  as a
dynamically generated resonance.~\cite{pelaez,eef} Dispersive
analyses are also made trying to understand the nature of $\sigma$
from $\gamma\gamma\to \pi\pi$ processes.~\cite{penn}

A study on the property of  the $\sigma$ meson using lagrangian
models encounter a problem: the large widths of the $\sigma$
pole~\cite{dis2} and the $\kappa$ pole~\cite{kappa,kappa2} indicate
that there must be very strong interactions involved. Since we know
little about how to do a calculation based on a lagrangian except
making perturbation expansions, certain unitarization approximation
has to be used.

An examination to the unitarized chiral perturbative amplitudes
finds a light and broad pole on the complex $s$ plane, in the
I,J=0,0 channel of $\pi\pi$ scattering, which is identified as the
$\sigma$ pole. It is  found that the $N_c$ trajectory of the
$\sigma$ pole has a non-typical behavior as comparing with that of a
normal resonance, e.g., a $\rho$ pole. Hence it is argued that the
$\sigma$ is a dynamically generated resonance from a lagrangian
without the $\sigma$ degree of freedom.~\cite{pelaez} This idea has
been carefully examined\cite{sun,guo}, and it is found that the
[1,1] Pad\'e approximation leads to a `$\sigma$' pole falling back
to the real $s$ axis in the large $N_c$ limit.  A correct
understanding on what does the [1,1] Pad\'e approximant mean is
obtained through these studies. The $\sigma$ pole position can be
obtained analytically from [1,1] Pad\'e approximant in the large
$N_c$ and chiral limit. Meanwhile the same pole position can be
found in a model $independent$ way, but under two additional
assumptions: 1) $s$-channel $\sigma$ pole dominance, 2) neglecting
the left hand cut.~\cite{sun}
 We will carefully study what does the second
assumption mean a while later in this talk. It is also pointed out
that the bent structure of the $\sigma$ pole trajectory with respect
to $N_c$ found in [1,1] Pad\'e approximant is in qualitative
agreement with what one finds in $O(N_f)$ $\sigma$ model, hence
suggesting a fundamental  role of the light and broad resonance pole
played at lagrangian level, even though it can be generated from
certain dynamical approximations.~\cite{guo} Here $N_f$ means the
number of flavors and one need not worry the lost of asymptotic
freedom when $N_f$ goes large. The real meaning of $O(N_f)$ toy
model is that, a large $N_f$ enables one to neglect the cross
channel dynamics legally and hence   solve the model analytically.

One constructs, using analyticity, unitarity, and $partial$ $wave$
dispersion relations, a factorized form for elastic scattering $S$
matrix element:~\cite{kappa}
\begin{equation}\label{Separable}
 S^{phy.}=\prod_iS^{R_i}\cdot S^{cut}\ ,
\end{equation}
where $S^{R_i}$ denotes the $i$-th $second$ sheet pole contribution
and $S^{cut}$ denotes the contribution from cuts or background. The
information from higher sheet poles is hidden in the right hand
integral which consists of one part of the total background
contribution,
\begin{eqnarray}\label{fs}
 S^{cut}&=&e^{2i\rho f(s)}\ ,\nonumber\\
 f(s)&=&\frac{s}{\pi}\int_{L}\frac{{\rm
 Im}_Lf(s')}{s'(s'-s)}+\frac{s}{\pi}\int_{R}\frac{{\rm
 Im}_{R}f(s')}{s'(s'-s)}\ .
\end{eqnarray}
The `left hand' cut $L=(-\infty, 0]$ for equal mass scatterings and
may have a rather complicated structure for unequal mass
scatterings. The right hand cut $R$ starts from first $in$elastic
threshold to positive infinity.  Estimates on the `left' cuts in
various channels of $\pi\pi$ and $\pi K$ scatterings using $\chi$PT
reveal a common feature: all the left cut contributions as defined
in Eq.~(\ref{fs}) are numerically found to be negative! This fact is
actually crucial to establish the existence of the $\sigma$ and
$\kappa$ pole in the present approach and also helps greatly in
stabilizing the pole location in the data fit. It is interesting to
notice that, there actually exists a correspondence of
Eq.~(\ref{Separable}) in quantum mechanical scattering theory,
obtained sixty years ago:\cite{NingHu}
 \begin{equation}\label{NHu}
S(k)=e^{-2ikR}\prod^{\infty}_{1}\frac{k_n+k}{k_n-k}\ ,
\end{equation}
 where $k$ is the (single) channel momentum and $k_n$ pole locations in the
 complex $k$ plane. The above formula is written down for any finite range
 potential, in $s$ wave. Notice that the
 Eq.~(\ref{NHu}) automatically predicts a negative background
 contribution!

Near physical threshold, on the left hand side of
Eq.~(\ref{Separable}), one can replace the physical amplitude by the
one calculated using $\chi$PT upto certain powers of external
momentum. One the right hand side one has a genuine parametrization
form for resonance contributions. If expand  both sides at, for
example, $\pi\pi$ scattering threshold, then one should be able to
obtain useful relations between low energy constants of $\chi$PT and
resonance parameters, without relying on any lagrangian models
describing these resonances. There is however a difficulty in
estimating the left hand cut integral on the $r.h.s.$ of
Eq.~(\ref{Separable}), in which both crossed channel $\pi\pi$ cut
and resonance exchanges contribute. Fortunately in the large $N_c$
limit the cut integrals can all be solved  analytically, and one
demonstrates that in the large $N_c$ limit the Eq.~(\ref{Separable})
is equivalent to the expression of partial wave dispersion
relation.~\cite{cillero} Useful relations can hence be obtained in
this way as expected.

This matching at $\pi\pi$ threshold has been done at $O(p^4)$ level
in Ref.~\cite{cillero}, and extended to $O(p^6)$ in
Ref.~\cite{cillero2}. Table~1 shows the matching results in
IJ=00,11,20, three different channels, at $O(p^2)$ level.
\begin{table}[!t]
\begin{tabular}{|c|c|c|c|c|}
  \hline
\rule[-0.7em]{0em}{1.9em}
 IJ  & $T(0)$ & $t_0^{\rm tR}$ & $t_0^{\rm sR}$ & $t_0^{\chi PT}$ \\
  \hline
\rule[-0.7em]{0em}{1.9em}
  $ 11$ & $-\frac{1}{24\pi f^2}$ & $\frac{4\Gamma_S}{9M_S^3}+\frac{2\Gamma_V}{M_V^3}$
        & $\frac{4\Gamma_V}{M_V^3}$ & 0 \\
  \hline
\rule[-0.7em]{0em}{1.9em}
  $ 00$ & $-\frac{1}{32\pi f^2}$ & $-\frac{4\Gamma_S}{3M_S^3}+\frac{36\Gamma_V}{M_V^3}$
        & $\frac{4\Gamma_S}{M_S^3}$ & $\frac{7}{32\pi f^2}$\\
  \hline
\rule[-0.7em]{0em}{1.9em}
  $ 20$ & $\frac{1}{16\pi f^2}$ & $-\frac{4\Gamma_S}{3M_S^3}-\frac{18\Gamma_V}{M_V^3}$
        & 0 & $-\frac{1}{16\pi f^2}$ \\
  \hline
\end{tabular}
\caption{{\small Summary of the different contributions  $T(0)$,
$t_0^{\rm tR}$, $t_0^{\rm SR}$ to the scattering lengths at leading
order in the $m_\pi^2$ expansion. $M$ and $\Gamma$ represent the
mass and width of a resonance, respectively. The unit of each
amplitude is in $m_\pi^2$.}} \label{tab.KSRF}
\end{table}
In table~1 $T(0)$ takes the $\chi$PT value at $s=0$, $t_0^{\rm tR}$
represents the resonance contribution in the crossed channel,
$t_0^{\rm sR}$ means the resonance contribution in the $s$ channel.
The sum of these three quantities equals to  $t_0^{\chi PT}$
denoting the $\chi$PT result on scattering amplitude at threshold.
It is realized that the sum leads to the same equation in three
different channels:
\begin{equation}
\label{sumksrf} \frac{1}{16 \pi f^2}=\frac{9\Gamma^{(0)}_V}{M_{\rm
V}^{(0)\, 3}}+\frac{2\Gamma^{(0)}_S}{3M_{\rm S}^{(0)\, 3}},
\end{equation}
where superscript $(0)$ means value in the chiral limit. The above
equation is some times called the generalized KSRF
relation.~\cite{igi} We learned an important lesson from the above
discussion, that is \textit{partial wave amplitudes remember
crossing symmetry}. If the crossed channel effects (i.e., the second
column in Table~1) were omitted incorrectly, then we would obtain
three different relations. It was stated earlier  that the [1,1]
Pad\'e approximation is equivalent to neglecting the crossed channel
cut in the large $N_c$ and chiral limit, hence it violates crossing
symmetry.

The above discussion also helps in understanding why the Pad\'e
approximation works good in the IJ=11 channel: because the cut
effects are very small in this channel, this property is sometimes
named as vector meson dominance in the literature; why works not so
good in the IJ=00 channel: because the left hand cut is
non-negligible in this channel; and why it runs into disaster in the
exotic IJ=20 channel: since in this channel there is only crossed
channel effects! One may draw a further conclusion: the quality of
Pad\'e approximation relies on whether in the corresponding channel
there is a single pole dominance.

If the $\sigma$ pole does not maintain a proper $N_c$ behavior,
i.e., $M\sim O(1)$, $\Gamma\sim O(1/N_c)$, then it would be very
difficult for the three channels to fulfil the constraints from
crossing symmetry simultaneously.

The Eq.~(\ref{sumksrf}) also explains, in one aspect, why the
$\sigma$ meson had been escaped from detection: it's contribution to
the $r.h.s.$ is numerically tiny as compared with that of the vector
meson. The vector meson plays a more prominent role than the scalar
meson in most cases.

As stated in the begining of this talk, if one construct a unitary
amplitude from perturbative amplitude calculated using chiral
perturbation theory lagrangian, one finds a light and broad pole on
the complex $s$ plane. Since this pole is absent in the original
chiral lagrangian,
 it is called `dynamically generated resonance'.~\cite{pelaez} It is
also found that the $N_c$ trajectory of such generated dynamical
$\sigma$ pole has a non-typical behavior as comparing with that of a
normal resonance, e.g., a $\rho$ pole. This is viewed as further
evidence to support the dynamical nature of the sigma pole. One may
even relate such exotic behavior to the tetra quark state.

Despite of the problem a unitarization approximation may encounter
as described in previously, the above phenomenon is itself
interesting and worthy of a careful examination.

To clarify the meaning of `dynamically generated resonance', we make
a simple exercise by calculating everything in the toy $O(N_f)$
linear sigma model. This model is exactly solvable, because when
$N_f$ is large Feynman diagrams that contribute are greatly reduced,
excluding those associated with crossed channel dynamics. One can
also pretend to solve it using [1,1] Pad\'e approximant, after
integrating out the sigma degree of freedom, at tree level. The
latter simulates the calculation in reality.

In both cases it is found that the $\sigma$ pole trajectory with
respect to $N_c$ in the $O(N_f)$ model behaves in an exotic way: it
does not like to fall down to the real axis but eventually it has
to.~\cite{xiao2}

Hence the bent structure of the sigma pole $N_c$ trajectory in
$O(N)$ model and its variation, similar to what is found in a more
realistic situation, does not at all suggest the $\sigma$ pole
itself is of dynamical origin.

We have shown the picture that a `dynamical resonance' generated
from a pure $\chi$PT lagrangian is neither necessary nor provides a
better understanding to the physics underlined.  We believe that the
$\sigma$ degree of freedom can be appropriately put in and described
by a lagrangian respecting chiral symmetry. Hence a lagrangian with
a linearly realized chiral symmetry is justified.

We have studied the scalar problem in the ENJ-L model (a model with
linearly realized chiral symmetry) in SU$_f$(3) content. The scalar
puzzle is of course not limited to the $\sigma$ only. One has to
examine the $\kappa$, $f_0(980)$, $a_0(980)$ altogether in order to
achieve a better understanding. Using a  crude K-matrix
unitarization scheme, we have shown that, the $\sigma$, $\kappa$ and
the $a_0(980)$ can be explained as members of one scalar
nonet.~\cite{su} Despite of the crude approximation being made, this
conclusion is not quite trivial and apparent. Remember that the
three scalars, not only have distorted masses, but also very
different widths. Also interestingly we found that the $f_0(980)$ is
not able to be included in the same nonet. The latter does not
disprove our picture that the $\sigma$ is accompanied by a whole
octet, since the $f_0(980)$ very likely owns a quite different
physical explanation.~\cite{baru,mark}

To conclude, we have found that all evidence accumulated so far are
consistent with the picture that the $f_0(600)$ resonance is nothing
but the $\sigma$ meson responsible for a spontaneous breaking of
linearly realized chiral symmetry, and be the chiral partner of the
Nambu--Goldstone bosons.

 {\bf Acknowledgement:}   The author  would like to
thank the organizers of the workshop. especially George Rupp, for
kind invitation to this nice workshop. This work is supported in
part by National Natural Science Foundation of China under contract
number
 10575002.

\end{document}